\documentstyle{article}
\input epsf

\newtheorem{theorem}{Theorem}
\newcommand{\proof}{\noindent {\bf Proof:\ \ }}
\newcommand{\qed}{\mbox{$\Box$}}

\begin{document}
\bibliographystyle{alpha}
\newcommand{\tetra}[4]{\mbox{$\fbox{\tiny
$\begin{array}{ccc}
\ \ \ \  & #1 & \ \ \ \ \\ 
   #4 & \ \ \ \    & #2 \\ 
         & #3 
\end{array}$}$}}

\title{TETRAVEX is NP-complete}
\author{Yasuhiko Takenaga  \\
Department of Computer Science \\
The University of Electro-Communications \\
Tokyo, Japan \\
takenaga@cs.uec.ac.jp 
\and
Toby Walsh\\National ICT Australia and 
University of NSW\\
Sydney, Australia\\
tw@cse.unsw.edu.au
}

\maketitle

\begin{quote}
\end{quote}

{\sc Tetravex} is a widely played one person computer
game in which you are given
$n^2$ unit tiles, each edge of which is labelled with
a number. The objective is to place each tile
within a $n$ by $n$ square such that all neighbouring
edges are labelled with an identical number. Unfortunately,
playing {\sc Tetravex} is computationally hard. More precisely, 
we prove that deciding if there
is a tiling of the {\sc Tetravex} board given $n^2$ unit tiles
is NP-complete. Deciding where to place the tiles
is therefore NP-hard. This may help to explain why
{\sc Tetravex} is a good puzzle.
This result compliments a number of similar
results for one person games involving tiling. 
For example, NP-completeness results have
been show for: the offline version of Tetris \cite{demaine1},
KPlumber (which involves rotating
tiles containing drawings of pipes to make a connected
network) \cite{kral1}, and
shortest sliding puzzle problems \cite{ratner1}.
It raises a number of open questions. 
For example, is the infinite version Turing-complete?
How do we generate {\sc Tetravex} problems which are truly 
puzzling as random NP-complete problems are often
surprising easy to solve? Can we observe 
phase transition behaviour? What about the complexity of the
problem when it is guaranteed to have an unique solution?
How do we generate puzzles with unique solutions?

\begin{theorem}
{\sc Tetravex} is NP-complete.
\end{theorem}
\proof 
Clearly it is in NP. Given a solution, we can
check it in polynomial time. To show completeness,
we use a reduction from 1in3-{\sc Sat} on purely
positive clauses. We will map a problem in $n$
variables and $m$ clauses onto a rectangular
{\sc Tetravex} problem of size $O(n)$ by $O(m)$. 
We can always convert a rectangular {\sc Tetravex}
problem into an essentially equivalent but larger square 
problem by adding suitable tiles.

There are five types of component used in the construction:
a horizontal assignment component along the top edge, 
vertical clause components, vertical and horizontal wiring tiles and 
and junction components to connect vertical to horizontal wires.
We assume the variables are labelled from 1 to $n$. 
The $i$th part of the assignment component consists of
four tiles that are in one of two configurations:
\begin{eqnarray*}
i={\it true} & : & 
\tetra{top}{i}{0}{i-1} \ \ 
\tetra{top}{i}{i}{i} \ \ 
\tetra{top}{i}{-i}{i} \ \ 
\tetra{top}{i+1}{0}{i} \ \ 
\\
i={\it false} & : & 
\tetra{top}{i}{0}{i-1} \ \ 
\tetra{top}{i}{-i}{i} \ \ 
\tetra{top}{i}{i}{i} \ \ 
\tetra{top}{i+1}{0}{i} 
\end{eqnarray*}
The value $top$ is used to ensure that these
tiles can only be placed along the top of
puzzle. No tile in the puzzle 
has $top$ at its bottom label. 
Actually, this is not essential and
we can label the top of these tiles 0.
However, it makes the proof easier if we
force the assignment component to
be on the top edge of the puzzle. 
The $i-1$ value in the leftmost tile,
and the $i+1$ value in the rightmost tile
are used to ensure that the assignment components
are laid out in order from left to right along
the top row of the puzzle. 
The value $0$ is used for internal tiles which
are not components.
We also start and end the top row with the tiles:
\begin{eqnarray*}
\tetra{top}{0}{0}{left} 
& \ \ \ldots \ \ &
\tetra{top}{right}{0}{n+1} 
\end{eqnarray*}
The value $left$ and $right$
are used to ensure
that a tile appears on the left or right
edge of the puzzle. No tile has
$left$ as its right label.
Similarly, no tile has
$right$ as its left label. 
This is not essential and we
could label them zero, but it again
makes the proof easier. 

This ``signal'' ($i$, $-i$ which is interpreted
as $i$ is $true$, 
or $-i$, $i$ which is interpreted as $i$
is {\it false}) is then
transmitted to the vertical clause components via
``wires''. There are vertical wires, horizontal wires and
junctions. A vertical wire is of the form:
\begin{eqnarray*}
\tetra{i}{X}{i}{X} 
& \ \ {\rm and} \ \ &
\tetra{-i}{X}{-i}{X} 
\end{eqnarray*}
Where $X$ is either 0 (if the wire is passing a blank part of
the puzzle), or the value of a horizontal wire being crossed. 
It is important to note that this value for $X$ is not
equal to $i$. When we cross a wire carrying the signal
from the $i$th variable, we use a junction component. 
These vertical wiring tiles can appear in either order depending
on the polarity of the signal being transmitted. 
A horizontal wire is of the form:
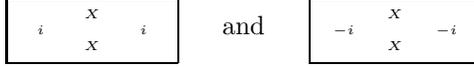
\begin{eqnarray*}
\tetra{X}{i}{X}{i} 
& \ \ {\rm and} \ \ &
\tetra{X}{-i}{X}{-i} 
\end{eqnarray*}
Where $X$ is again either 0 (if the wire is passing a blank part of
the puzzle), or the value of a vertical wire being crossed. 
We again note that this value for $X$ is not
equal to $i$. These horizontal wiring
tiles can appear in either order depending
on the polarity of the signal being transmitted. 

A junction connects a vertical pair of wires
with a horizontal pair of wires. Each junction
is labelled with an unique number $j$ where $j>n+m$. 
The junction consists of one of two possible 
arrangements of four tiles:
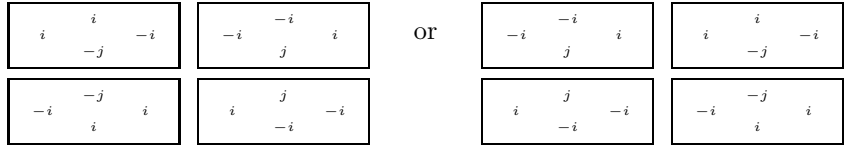
\begin{eqnarray*}
\tetra{i}{-i}{-j}{i} \ \ \tetra{-i}{i}{j}{-i} 
& \ \ {\rm or} \ \ &
\tetra{-i}{i}{j}{-i} \ \ \tetra{i}{-i}{-j}{i} \\
\tetra{-j}{i}{i}{-i} \ \ \tetra{j}{-i}{-i}{i} 
& & 
\tetra{j}{-i}{-i}{i} \ \ \tetra{-j}{i}{i}{-i} 
\end{eqnarray*}
The junction turns the vertical signal $i$, $-i$ into
the horizontal signal $\begin{array}{c} i \\ -i\end{array}$,
or the vertical signal $-i$, $i$ into
the horizontal signal $\begin{array}{c} -i \\ i\end{array}$.

Whilst it is possible to stack two wiring
tiles horizontally and place them between
the left and right half of the junction component,
we cannot then line up with the wiring components
coming down from the assignment component. 
Similarly, it is impossible to put two wiring
tiles between the top and bottom half of the
junction component. 
Therefore this junction component must
occur in this two by two form. 

Finally, there is the clause component. For each
clause, we have 12 vertically arranged tiles.
Suppose the $p$th clause is $i \vee j \vee k$. 
We label this with an unique number, $c=n+p$. 
The clause component consists of a top tile, a buffer tile, then 
two tiles connected to the wires bringing
in the $i$, $-i$ or $-i$, $i$ signal,
another buffer tile, 
two tiles connected to the wires bringing
in the $j$, $-j$ or $-j$, $j$ signal,
another buffer tile, 
two tiles connected to the wires bringing
in the $k$, $-k$ or $-k$, $k$ signal,
one more buffer tile, 
and finally the bottom tile. 
There are thus four buffer tiles in total. Two 
of them are labelled: 
$$ \tetra{-c}{0}{-c}{left}$$
The $left$ label ensures that this tile is
placed along the left edge of the puzzle.
Such tiles will be
adjacent to any pair of wires bringing
in a {\it false} signal. 
The other two buffer tiles, which are immediately
above and below the pair of 
wires bringing in the $true$ signal, are: 
\begin{eqnarray*}
\tetra{-c}{0}{c}{left}
& \ \ {\rm and} \ \ &
\tetra{c}{0}{-c}{left}
\end{eqnarray*}
The top and bottom tiles of the clause component are:
\begin{eqnarray*}
\tetra{0}{0}{-c}{left}
& \ \ {\rm and} \ \ &
\tetra{-c}{0}{0}{left}
\end{eqnarray*}
Finally, the two tiles connecting the wires bringing
in the $i$ signal are:
\begin{eqnarray*}
\tetra{c}{i}{-c}{left}
& \ \ {\rm and} \ \ &
\tetra{-c}{-i}{c}{left}
\end{eqnarray*}
A similar pair of tiles connect the wires
bringing in the $j$, $-j$ or $-j$, $j$ signal,
and the $k$, $-k$ or $-k$, $k$ signal. 

To illustrate the clause component, we give the 12 tiles
representing the clause $i \vee j \vee k$, with $i$ being
the only variable set true. For brevity, we lay the tiles
out horizontally in four (connected) rows:
\begin{eqnarray*}
\tetra{0}{-c}{left}{0} \ 
\tetra{0}{c}{left}{-c} \ 
\tetra{i}{-c}{left}{c} \ 
\tetra{-i}{c}{left}{-c} \ldots \\
\ldots \tetra{0}{-c}{left}{c} \
\tetra{-j}{c}{left}{-c} \
\tetra{j}{-c}{left}{c} 
\ldots \\
\ldots 
\tetra{0}{-c}{left}{-c} \
\tetra{-k}{c}{left}{-c} \ 
\tetra{k}{-c}{left}{c} \ldots \\
\ldots 
\tetra{0}{-c}{left}{-c} \
\tetra{0}{0}{left}{-c} 
\end{eqnarray*}
With the clause component, we pick out one (and only one) 
of the three input pairs of wires to be $true$. 
Since the top-most and bottom-most labels
of each clause component is zero, we can lay
out the clause components in any order. However,
they must be along the left edge of the puzzle
as the left side of each tile is labelled
$left$ and no tile has $left$ as its righthand
label. 

Finally, we fill in the rest of the puzzle
with tiles labelled just zero. 
The puzzle is $4n+2$ tiles wide. In the
assignment component along the top edge,
each of the $n$ variables is assigned a value using a block of 4 tiles.
We also have the start and end of row tiles. 
This makes $4n+2$ in total. 
The puzzle is $12m+1$ tiles high. Each of the $m$
clauses using a component with 12 tiles. There is
also the top left tile which starts the assignment
component. 
This makes $12m+1$ in total. 
Within the puzzle, there are $3m$ junction components.
Each clause requires 3 junctions, one for each
variable. There are $24nm-12m$ vertical wiring tiles. 
Each of the $n$ variables has a vertical wire
running the $12m$ tiles from the assignment component
at the top of the puzzle to the bottom edge of the puzzle. 
Each of these wires is two tiles wide. This gives $24nm$ tiles.
There are, however, no wiring tiles where we have 
junction components. There
are $3m$ junctions in total, each consisting of 4 tiles.
Hence, there are $24nm-12m$ vertical wiring tiles in total. 
Finally, there are $24nm-6m$ horizontal wiring tiles. 
Each of the $m$ clauses has three horizontal wires
running the $4n+1$ tiles from the clause
component at the left edge of the puzzle to the
right edge of the puzzle. Each of these wires is
two tiles wide. This gives $6m(4n+1)$ tiles.
There are, however, no wiring tiles where we have 
junction components. There
are $12m$ such tiles. Hence, 
there are $24nm-6m$ horizontal wiring tiles in total. 

Suppose there is an assignment which satisfies
just one variable in each clause. Then it is 
easy to see that there is a proper tiling of the puzzle.
Suppose, on the other hand, that there is no
such assignment. Assume there was a proper tiling
of the puzzle. This means that there must be 
an assignment component along the top edge.
Now the vertical wires can only fit in the puzzle if
they are connected to this component. 
Similarly, there must be a clause component 
along the left edge. 
The horizontal wires can only fit in the puzzle if
they are connected to this component. 
Finally, the junction components
can only fit into the puzzle if they wire
up the horizontal and vertical wires correctly. 
Thus we have a correct wiring of the circuit. However,
this is only possible if we can satisfy the 1in3-{\sc Sat}
problem. Hence, the 1in3-{\sc Sat} problem
is satisfiable iff there is a proper tiling of this
puzzle. 
\qed

Some observations can be made about this result.
We can add
the circular boundary condition that the top
edge of the square matches
the bottom, and the left edge matches the right. 
{\sc Tetravex} with such boundary
conditions remains NP-complete
(since it is easy to modify the reduction so all
edges are labelled zero). 
We can also generalize {\sc Tetravex} to 3 (or more) dimensions. 
The problem remains NP-complete as we need only one plane
for the reduction and can use (hyper)cubes labelled with
zero everywhere else. 
Finally, our reduction requires $O(n+m)$ integer labels. 
It is an open question if the problem remains
NP-complete when we have just $O(1)$ different labels.

\newcommand{\etalchar}[1]{$^{#1}$}

\end{document}